\documentclass[a4paper,12pt]{article}

\usepackage{geometry}
\usepackage[pdftex]{graphicx}
\usepackage{setspace} 
\usepackage{amsmath}
\usepackage{empheq} 
\usepackage{comment} 
\usepackage{multicol} 
\usepackage{enumitem} 
\usepackage[justification=centering]{caption}
\usepackage{caption}
\usepackage[table]{xcolor} 
\usepackage{multirow} 
\usepackage{float}
\usepackage{verbatim}
\usepackage{url}
\usepackage{appendix}
\usepackage{ulem}
\usepackage[hang]{footmisc}

\usepackage[english]{babel}
\usepackage{textcomp}
\usepackage{amssymb}
\usepackage{booktabs}
\usepackage{adjustbox}
\usepackage{booktabs}
\usepackage{rotating}
\usepackage{tablefootnote}
\usepackage[flushleft]{threeparttable}
\usepackage{tabularx}
\usepackage{natbib}
\usepackage{titlesec}
\usepackage{xr-hyper}
\usepackage[hidelinks]{hyperref}
\usepackage{pdflscape}
\usepackage{rotating}
\usepackage{anysize}
\usepackage{titlesec}

\marginsize{2.0 cm}{2.0 cm}{2.0 cm}{2.0 cm}

\bibpunct{(}{)}{;}{a}{,}{,}

\begin{document}
	
	\pagestyle{plain}
	
	\title{\textsc{School-age Vaccination, School Openings and Covid-19 diffusion}}
		
	\author{\textbf{Emanuele Amodio}\thanks{Department of Health, Promotion, Mother and Child Care, Internal Medicine and Medical Specialties, University of Palermo, Piazza delle Cliniche, 2, 90127 Palermo, Italy, e-mail: emanuele.amodio@unipa.it.}\\  University of Palermo 
		\and 
		\textbf{Michele Battisti}\thanks{Department of Law, University of Palermo, Piazza Bologni 8, 90134 Palermo, Italy, e-mail: michele.battisti@unipa.it.}\\  University of Palermo, ICEA 
		\and 
		\textbf{Antonio Francesco Gravina}\thanks{Department of Law, University of Palermo, Piazza Bologni 8, 90134 Palermo, Italy, e-mail: antoniofrancesco.gravina@unipa.it.}\\  University of Palermo
		\and 
		\textbf{Andrea Mario Lavezzi}\thanks{Department of Law, University of Palermo, Piazza Bologni 8, 90134 Palermo, Italy, e-mail: andreamario.lavezzi@unipa.it.}\\  University of Palermo, ICEA
		\and 
		\textbf{Giuseppe Maggio}\thanks{Department of Law, University of Palermo, Piazza Bologni 8, 90134 Palermo, Italy, e-mail: giuseppe.maggio12@unipa.it.}\\  University of Palermo
	}
	\maketitle
	
\vspace{-1cm}
\begin{abstract}
	\noindent Do school openings trigger Covid-19 diffusion when school-age vaccination is available? We investigate this question using a unique geo-referenced high frequency database on school openings, vaccinations, and Covid-19 cases from the Italian region of Sicily. The analysis focuses on the change of Covid-19 diffusion after school opening in a homogeneous geographical territory. The identification of causal effects derives from a comparison of the change in cases before and after school opening in 2020/21, when vaccination was not available, and in 2021/22, when the vaccination campaign targeted individuals of age 12-19 and above 19. The results indicate that, while school opening determined an increase in the growth rate of Covid-19 cases in 2020/2021, this effect has been substantially reduced by school-age vaccination in 2021/2022. In particular, we find that an increase of approximately 10\% in the vaccination rate of school-age population reduces the growth rate of Covid-19 cases after school opening by approximately 1.4\%. In addition, a counterfactual simulation suggests that a permanent no vaccination scenario would have implied an increase of 19\% in ICU beds occupancy.
\end{abstract}

	\vspace{0.2 cm}
	
	\pagenumbering{arabic}
	\doublespacing \setcounter{page}{1}
	
\section{Introduction\label{se:intro}}
	
The Covid-19 pandemic determined the flourishing of a substantial amount of studies addressing the health and socio-economic determinants of its diffusion, aiming also at identifying measures tied to contain its direct and indirect costs for the society.

Schools, and the students' population, have been a crucial aspect of the discussion for at least two reasons. On the one hand, among the restrictive policies implemented during the first and second wave of the pandemic, when anti-Covid-19 vaccines were not available, school closure has been a measure widely adopted, together with more general measures of lockdown of economic activities. Closing the schools, and implementing distance learning, was based on the assumption that the interactions implied by attending schools might have been an important driver of the spread of Covid-19 in the population. The recent literature addressing the issue, often exploiting school openings for identification purposes, found mixed results (see \citealp{svaleryd2022covid}, for a detailed survey). On the other hand, closing the schools raised concerns on the costs in terms of lost opportunities of accumulating human capital, as well as on the psychological costs of the students and distress of the students' families (see \citealp{stantcheva2022inequalities}, and references therein).


In December 2020, following the approval by the Food and Drug Administration in the US and the European Medicines Agency, anti-Covid 19 vaccines became available and recommended for individuals older than 16 years, and were subsequently approved in May 2021 for adolescents in the age bracket 12-15. In this new context, some crucial questions naturally arise: when vaccination is available to the students' population, do school openings still represent a potential triggering factor of Covid-19 diffusion? Is school closure, therefore, still to be recommended as an effective mitigation policy?

In this article we try to answer these questions by analyzing  granular data from the Italian region of Sicily, comparing the effect of school openings during the school year 2020/21, when vaccines were not available, to the effect in 2021/22, when vaccines were available for the students' population and for the population at large. Considering granular data from a homogeneous territory and the same period of the year may help to account for a wide range of social and institutional confounding factors and for the effects of seasonality. 

Our analysis relies on a dataset obtained by merging geo-localized data on Covid-19 cases, information on age vaccination exposure, schools' geographical location and school opening dates. In particular, we build an indicator of local (i.e. at census micro-area level) vaccine exposure from detailed data on daily vaccinations by age at municipal level, and on the demographic structure of the census area population.

Our results suggest that school-age vaccination played a major role in reducing Covid-19 cases diffusion after school openings in Sicily. Specifically, we show that school openings in 2021 is associated to a differential impact when compared to 2020, with almost no effects on Covid-19 diffusion at the local level. The positive effect of school openings on Covid-19 diffusion, identified by \citet{amodio2022schools}, appears to be fully mitigated by vaccination, with higher effects of vaccination in areas where the share of vaccinated school age population is lower and in areas with lower population density (that in the literature is often associated with higher and denser social interactions as we discuss below). In addition, we show by a counterfactual analysis that the diffusion of vaccination is associated to a reduction of hospitalizations in ICU of approximately 19\%.

The work is organized as follows. Section \ref{se:literature} provides a review of the relevant literature; Section \ref{background} describes Covid-19 transmission and vaccinations in Sicily, while Section \ref{se:data} introduces the dataset. Section \ref{se:strategy} specifies the econometric models we utilize. Section \ref{se:econometrics} presents the main results, a set of robustness checks and a heterogeneity analysis; Section \ref{se:conclusions} concludes.


\section{School Openings, School Closures, Vaccinations and Covid-19 Diffusion: a Review of the Literature}\label{se:literature}



Our contribution speaks to three related strands of literature. The first one is about the effects of school openings on Covid-19 diffusion. The second refers to the efficacy of school closures as a mitigation policy. The third relates to the overall effect of vaccination in containing the epidemic. 

The effects of school opening as a trigger of Covid-19 diffusion is addressed by several studies. Among these, works such as \citet{amodio2022schools}, \citet{Chernozhukove2103420118}, \citet{vlachos2021effects} find a positive effect of school openings on the growth rate of Covid-19 diffusion in a range of 2\%-5\%. Differently, other studies find nonsignificant or mixed effects. In particular, \citet{isphording2021does} analyze the German case taking into account the school openings after the summer break of 2020, and do not find a positive impact on Covid-19 diffusion.  \cite{yoon2020stepwise} focus on the first wave of Covid-19 in Korea and investigate the step-wise transition from school closure to opening, both online and offline, and find that offline schooling did not lead to any substantial outbreak among the youths. Other works, such as \cite{keeling2021impact} study eight different school reopening strategies implemented for primary and secondary schools in England. The authors find a high degree of heterogeneity, with half-sized classes or younger cohorts not associated to increased levels of infections, and the reductions in community social distancing exacerbating the impacts of school openings. Finally, \citet{oster2021Openings} find regional differences of school openings in the US, with no significant effects in most of US states and positive effects in Southern states, in a setting where different teaching methods (in-person, hybrid, remote) is controlled for. Our paper, however, is the first one that compares the effect of school openings on Covid-19 diffusion in a period with no vaccines available to a period in which vaccines were introduced and gradually extended to different age cohorts of the population, focusing on the effect of vaccination.\footnote{A partial exception is \citet{isphording2021schools}, who also analyze a period in which vaccination was available in Germany. However, they do not focus on the effects of vaccinations but on the effect of mandatory testing in German schools, concluding that this can be an effective policy in containing the infection while schools are open.}

The effect of school closures as a mitigating policy, especially during the first wave of Covid-19 diffusion is analyzed, for example, by \citet{flaxman2020estimating} who find a general, positive effect of lockdowns on Covid-19 transmission, but cannot identify a specific effect of school closures. This depends on the fact that they rely on a cross-country study which does not allow to disentangle the effect of general lockdowns from the one of school closures. Differently, \citet{alfano2022effects} implement a panel data analysis of the effect of school closures in a sample of European countries. \citet{alfano2022effects} finds that, given a time lag of 10 to 40 days, school closures have a sizable, negative effect on Covid-19 diffusion, measured by daily cases of infections. \citet{ferguson2020impact} study the case of mitigation policies in UK, and estimate a specific effect of school closures. They find that school closures have a sizable effect in reducing Covid-19 diffusion, estimated as a reduction of 14\% of the peak demand of ICU beds. Differently, \cite{fukumoto2021no} make use of municipality level data to investigate the effects of schools' closures in Japan during spring 2020 but find no causal links of school closures in reducing the spread of COVID-19. These studies, however, refer to a period in which vaccinations were not available and, therefore, conclusions based on their results on the appropriateness of school closures as a mitigating policy should be re-evaluated at the time in which vaccines became available. In addition, none of these studies utilizes granular data as in this work, which allow to specifically focus on the school-age population while accounting for a number of confounding factors and, therefore, to formulate a more accurate evaluation of school closures as a mitigation policy. This is all the more important given that, as mentioned in Section \ref{se:intro}, school closures imply a host of negative effects, such as losses in human capital accumulation and earnings over the lifetime (see, e.g., \citealp{agostinelli2022great, psacharopoulos2021covid,fuchs2020long}, and  \citealp{stantcheva2022inequalities}, for an updated account of the literature on this point).\footnote{Other studies considered different mitigation policies related to schools, such as social distancing. See, e.g., \citet{oster2021socialDist}.}



Finally, a growing literature started to analyze the effect of vaccinations on Covid-19 diffusion. For example, the extensive study of \cite{kim2022vaccination} shows that vaccination reduced diffusion in the medium run, together with other non-pharmaceutical interventions. 
Among others, \cite{heath2021safety} study a set of more than 15,000 individuals who underwent randomization for Covid-19 vaccine, and find that the level of efficancy of vaccination efficacy of is about 86.3 percent against the alpha variant, while raises up to 96.4\% when considering the original variant. \cite{francis2021review} conduct an extensive review of COVID-19 vaccine sub-types evidencing the associated efficacy and their geographical distribution, finding that this varies between 66.9\% for Janssen, widely diffused in North-America, South-America, South Africa and Europe, to 95\% for Pfizer which covers the same continents but is widely diffused also in Australia and middle-east.
Other studies addressed the strategies and the optimal level of vaccination to reach heard immunities and reduce contagion. For example, \citet{cot2021impact} focus on human mobility and Covid-19 vaccine to investigate the effect of the US vaccination strategies on the pandemic dynamics in 2020/2021. The authors find that vaccination alone may not impede the outbreak of large contagions and that social distancing measures may become necessary until an high level of immunity is achieved. \citet{coccia2022optimal} aims at identifying the optimal level of vaccination. expressed as number of doses per 100 inhabitants, for a sustained reduction in Covid-19 cases and deaths. The results of this study suggest this level is reached when at least 80 doses of vaccines per 100 inhabitants are administered. To the best of our knowledge, however, this is the first work that tries to assess the effectiveness of anti-Covid 19 vaccinations through a comparison of school openings at times  where vaccines were not or were available. Besides providing estimates of the effect of school openings on Covid-19 diffusion in these two periods, we will propose an estimation by a counterfactual analysis of the reduction in ICU hospitalizations due to school-age vaccinations. This may help to give a more variegated policy answer based on cost-benefit analysis than opening or closing schools (and implementation of other kind of restrictions), insofar as there exists the additional element of health and non-health interventions.

\section{Background: Covid-19 Diffusion and Vaccinations in Sicily}\label{background}

\begin{figure}[H]
	\centering
	\minipage{0.5\textwidth} 
	\includegraphics[width=1\linewidth]{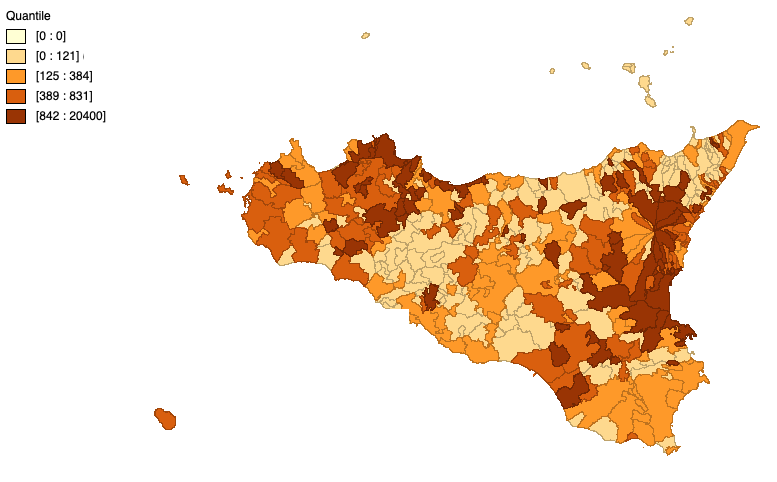}
	\endminipage\hfill
	\minipage{0.5\textwidth}
	\includegraphics[width=1\linewidth]{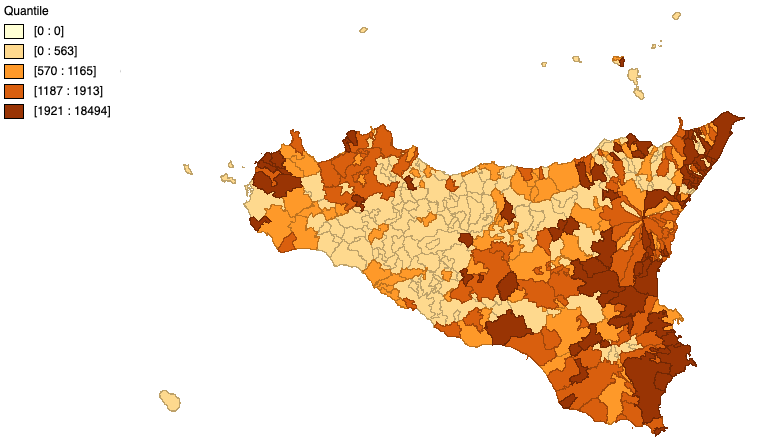}
	\endminipage\hfill
	\caption{Covid Cases by millions of inhabitants in 2020 and 2021}. 
	\label{Fig_Cases}
\end{figure}

Sicily and the South of Italy have been only marginally affected by the Covid-19 pandemic during the first wave, but the level of Covid-19 cases substantially increased from the second wave onwards. Figure \ref{Fig_Cases} displays the population-weighted cases (by quintiles) at the end of the summers of 2020 and 2021. While the number of cases are substantially higher in the second year (right panel), their spatial distribution follow similar paths, with more cases in more touristic areas, such as the North-West and East coasts, and less cases in the inner areas, which are less populated and often mountainous.




As in the rest of Italy, the vaccination campaign started in January 2021 for the adult population, and was progressively extended to younger cohorts, up to 2nd June 2021 when in Sicily (in Italy this possibility was open by May 29th) every citizen older than 11 was allowed to receive two doses of vaccine in a window of 28 days. From June 30, therefore, the population aged 12-19 had the possibility of being fully vaccinated, i.e. of receiving two doses of anti-Covid-19 vaccine. 



\begin{figure}[H]
	\centering
	\minipage{0.5\textwidth} 
	\includegraphics[width=1\linewidth]{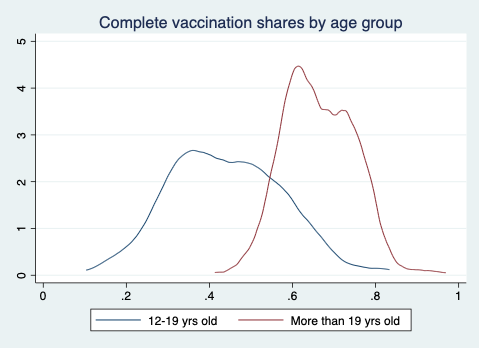}
	\endminipage\hfill
	\minipage{0.5\textwidth}
	\includegraphics[width=1\linewidth]{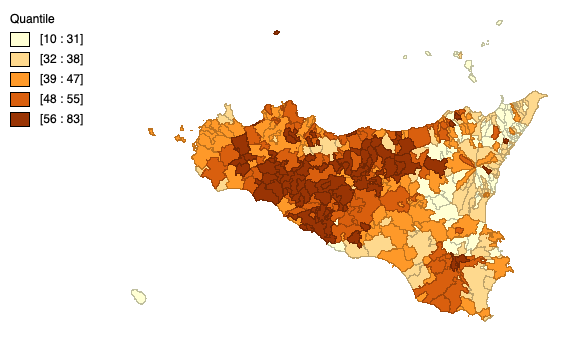}
	\endminipage\hfill
	\caption{Vaccination rates: densities for different demographic gropus (left panel), and distribution in the geographical space}
	\label{Fig_Vaccin}
\end{figure}

Figure \ref{Fig_Vaccin} (left panel), shows the density of shares of vaccinated population across municipalities for two groups at the time of school opening in 2021: the school-age population, i.e. of age 12-19,\footnote{By considering the age bracket 12-19, we are analyzing students attending compulsory school (\textit{scuola media}), up to age 13, and students attending high-school. The lower bound of this age bracket is given to the lowest age for which vaccination was available at the time of school openings in 2021, while the upper bound is given by the age at which students attend the last year of high-school. Vaccinations for pupils younger than 12, i.e. in the age bracket 5-11, were introduced in Italy only in December 2021.} and the population of age greater than 19. In particular, we see in Figure \ref{Fig_Vaccin} (left panel) that in the school-age population the average is lower, which is expected because of the different timing of the vaccinations, but the variance is much higher than in the older group.

Figure \ref{Fig_Vaccin} (right panel) indicates that vaccination rates for 12-19 years old display a wide heterogeneity across the geographical space, which adds up to the heterogeneity of Covid-19 cases in the geographical space highlighted in Figure \ref{Fig_Cases}.

Given these patterns of Covid-19 cases and vaccinations across space, we will consider the case in which Covid-19 contagion spreads locally and then has further spillovers. Specifically,  we will follow the strategy of \cite{amodio2022schools} and assume that contagion may be triggered by school openings in areas geographically close to the schools. In particular, we will assume that school openings affected the geographic micro-areas (i.e. the Italian census cells) within a ray of one km from the schools (in case of more than one school we will use a weighted average based on the number of students). In the next section we provide the details on the dataset.




%

\section{Data\label{se:data}}
The analysis relies on a set of data on 33,604 census cells observed four weeks before and four weeks after school openings in the school years 2020/21 and 2021/22. Census cells are micro-areas defined by the Italian National Statistical Office (ISTAT) to conduct the census of the population.\footnote{The last census in Italy was conducted in 2011. This represents the most recent census for which data are available.} Sicily counts 36,681 census cells with an average population of about 152 inhabitants and a median area of 0.13 km$^2$. Part of these cells do not have any resident population and thus are excluded from the analysis. The dependent variable in our econometric analysis is the weekly change in the log of Covid-19 cases at census cell level obtained from \cite{ISS:2020}, which provides information on new Covid-19 cases on a daily basis.\footnote{To include the zero-valued observations we add 1 before taking the log. This result is robust when using the Inverse Hyperbolic Transformation developed by \cite{Bellemare-et-al:2020}.} We aggregate Covid-19 cases at weekly level to account for the serial time of infection (that is the median time to develop infection, after contagion), following the standard in the literature (e.g., \citealp{cereda2020early}).
   
To capture the effect of school openings, we build a dummy activating when at least one school attended by students in the age bracket 12-19 opens within a ray of 1km from the centroid of a census area. This variable is equal to zero before the week in which school opened and takes on a value of one afterwards, in both school years 2020/21 and 2021/22. We focus only on the 4,223 public schools in Sicily, as these contain more than 95\% of students in the region, for grades where school is compulsory (in the periods considered Sicily counts approximately 790,000 students, of which 240,000 in high schools).

\defcitealias{ISS:2020}{ISS (2020)}
From \citetalias{ISS:2020} we also gather data on vaccinations, which includes information on the number of administered first and second doses by age. Differently from the Covid-19 cases, these data do not contain the indication of the census cell of residence of the vaccinated individual, so we can only utilize  data on vaccinations at municipal level. We use these data to develop a set of indicators on vaccination coverage by age. First, for consistency, we create a variable on the total number of vaccinated individuals by class age and municipality.
 Then, we combine the vaccination data with population data from ISTAT to build two indicators on the share of the 12-19 population and on the share of the population of 19 or older, who received the first and second dose in a given municipality. Finally, we combine vaccination information at municipality level with age-structure information at cell level from the 2011 Census to develop a cell-level indicator measuring exposure to vaccination in a census cell $i$ as follows:
\begin{equation}
	Exp\_Vaccines_{age, i} = \left( \frac{vaccines_{age, m}}{pop\_mun_{age,m}} \right) * \left( \frac{pop\_cens_{age, i}}{pop\_cens_i} \right), \label{eq:exposure}
\end{equation}
where $age = \lbrace 12-19, >19 \rbrace$; $vaccines_{age, m}$ represents the number of vaccinated individuals (with either one or two doses) in age group $age$ administered in Sicily at the date of September 16th 2021 in municipality $m$, and $pop\_mun_{age, m}$ indicates the population in age group $age$ in municipality $m$; $pop\_cens_{age, i}$ denotes the population in age group $age$ in the census area $i$, while $pop\_cens_i$ represents the total population of census cell $i$.\footnote{We exclude unreliable values for the share of 12-19 individuals (i.e. greater than 50\%), probably due to measurement errors, at the census area level. These account for less than 0.1\% of total observations. Results of the empirical analysis are consistent with the inclusion of these observations, and are available upon request.}

This logic of this variable is that the higher the share of vaccinated population in an age group in a municipality, and the higher the share of population of that age group in a census area, the more exposed the census area is to vaccination. The implicit assumption is that the within a city vaccination rates for an age group (i.e. 12-19) are quite similar across census cells.\footnote{Lacking census level vaccination data we may quote  anecdotal evidence supporting this assumption, especially in the age range 12-19 where the expected differences in terms of, e.g. education and habits, appear to be rather low. For instance \cite{tiu2021characterizing} find out that vaccination coverage is highly heterogeneous across the US territory and this determines the existence of  spatial clusters of undervaccination, mainly distributed in the southern states. On the other hand, only 3 municipalities out of 390 in our sample have more than 200,000 inhabitants, so that the room for high within-city heterogeneity appears quite limited. If we exclude these 3 municipalities, or all those with more than 100,000 inhabitants (losing respectively 15 and 18\% of observations) results are consistent with those presented below.} 

Finally, we collected data on the total number of beds in intensive care units (ICU) in the municipality's hospitals, and on the distance of the municipality centroid from the closest hospital with a ICU, gathered from data on the health structures from the Ministry of Health.\footnote{The data on health structures are available at the following link: \url{https://www.salute.gov.it/portale/documentazione/p6_2_8_1.jsp}.} For a municipality hosting at least one hospital with ICU, the distance indicator takes value zero. For a municipality without a hospital with ICU, the number of beds takes a value equal to zero, while the distance takes a real positive value.

Table \ref{tab_stats} reports the descriptive statistics of the variables, both at the municipality and census cell level, that we will use in the econometric analysis.\footnote{In Table \ref{tab_stats} by ``share" we refer to the number of vaccine doses divided by the population. For example, an individual that received two doses is counted twice in the calculation of the share. For this reason, the share referred to receiving at least one dose can be greater than one.}

\begin{table}[H]
	\singlespacing
	\captionsetup{format=plain, labelfont=bf}
	\caption{Summary statistics }
	\centering
	\footnotesize
	\begin{tabular}{lccccc}
		\toprule
		\multicolumn{6}{c}{Municipality level} \\
		\midrule
		& Obs    & Mean   & Std. Dev. & Min    & Max   \\
		\cline{2-6}
		Share of vaccinated (12-19) (2 doses)      & 390    & 0.44   & 0.14     & 0.10  & 0.83 \\
		Share of vaccinated (12-19) (at least 1 dose)       & 390    & 1.07   & 0.27     & 0.25   & 1.72 \\
		Share of vaccinated ($>$19) (2 doses) & 390 & 0.66 & 0.09 & 0.41 & 0.97 \\
		Share of vaccinated ($>$19) (at least 1 dose) & 390 & 1.40 & 0.16 & 0.88 & 2.07 \\
		ICU bed places &  390 & 1.03 &  7.66 & 0 & 123 \\
		\midrule
		\multicolumn{6}{c}{Census area level} \\
		\midrule
		Covid-19 cases, growth rate (2020)      & 134,416 & 0.02   & 0.24      & -3.26 & 3.95 \\
		Covid-19 cases, growth rate (2021)      & 100,812 & -0.01 & 0.24   & -3.14 & 2.40 \\
		Covid-19 cases, growth rate (2020-2021) & 33,585  & -0.03 & 0.11     & -1.02  & 0.79 \\
		Exposure to vaccination (12-19) (2 doses)  & 32,490  & 0.04  & 0.03     & 0      & 0.31 \\
		Exposure to vaccination (12-19) (at least 1 dose) & 32,490  & 0.09  & 0.06     & 0      & 0.73 \\
		Exposure to vaccination ($>$19) (2 doses) & 32,490 & 0.52 & 0.12 & 0 & 0.93 \\
		Exposure to vaccination ($>$19) (at least 1 dose) & 32,490 & 1.10 & 0.26 & 0 & 1.91 \\
		\bottomrule \label{tab_stats}
		
	\end{tabular}
\end{table}		

In the next section we describe our empirical strategy.

\section{Empirical strategy} \label{se:strategy}
The empirical strategy relies on three different specifications. The first specification captures the differential effects between school opening in 2020/21 and 2021/22. We do this by using a Diff-in-Diff (DiD) dynamic process model with fixed effects and week dummies, as in \cite{Chernozhukove2103420118} and \cite{amodio2022schools}.\footnote{With respect to \cite{amodio2022schools}, the time windows under consideration are shorter. This is done to avoid further confounding factors, such as the beginning of the colder season and the spread of the Omicron variant, which was observed from November 2021 in Italy.}


The empirical specification takes the following form:
\begin{equation}\hspace{-0.5cm}
	\footnotesize
	\Delta lnCovid_{i,t, y}=\alpha_i+ \rho \Delta  lnCovid_{i,t-1,y} + \sum_{j=3}^4 \beta_j  lnCovid_{i,t-j} + \lambda S_{i,t-2} +\eta S_{i,t-2} * Year_{2021/22} + \tau C_{i}+g T_{t}  +u_{i,t} 
	\label{eq:np_1}
\end{equation}

\hspace{25mm} with $\quad i=1,2,\ldots,n$; \hspace{2mm} $t=1,2,\cdots, T$; \hspace{2mm}  y= 2020/21, 2021/22; \\
where $\Delta lnCovid_{i,t,y}$ denotes the growth rate of Covid-19 cases in census area $i$ in week $t$ and school year $y$. The equation includes the lag of the dependent variable and the term $lnCovid_{i,t-j}$, denoting the natural logarithm of Covid-19 cases in the same census area $i$, measured at times $t-3$ and $t-4$, following an approach similar to \cite{Chernozhukove2103420118}, who show that this can be seen as an empirical specification derived from a theoretical SIR model (see also \citealp{amodio2022schools}). The main explanatory variable of interest is the coefficient $\lambda$ of the dummy $S_{i,t-2}$ on school openings, which takes value 0 in the four weeks before the opening, and 1 after the week of school opening. As in \cite{amodio2022schools}, this variable enters with a 2-week lag to account for the time to detect the contagion, defined also as the serial time of infection. The coefficient $\lambda$ captures the impact of school opening on Covid-19 diffusion and can be interpreted as a percentage increase in the growth rate. The term $\eta$ denotes the coefficient of the interaction between the school opening dummy $S_{i,t-2}$ and the year dummy $Year_{2021/22}$, and it aims at capturing the differential effect of school opening in the second year of the pandemic.\footnote{We do not include the $Year_{2021/22}$ dummy alone as this is collinear to the week dummies $T_{t}$.} The model includes census area fixed effects $C_{i}$, which account for short term time-invariant unobserved characteristics, such as the population profile or level of education, and week dummies $T_{t}$, which account for common shock in time, such as an increase in the number of Covid-tests available from a given week onwards. Finally, the term $u_{i,t}$ is a robust error term clustered at census area level.

In a second specification we study whether the differential increase in cases during the four weeks following school opening in 2021/22 and in 2020/21 can be attributed to school-age vaccinated population, and how this effect compares to the one associated to the vaccinated population of age higher than 19.\footnote{We refer to vaccinated population/individuals as the one(s) who received two doses of Covid-19 vaccine. To test this assumption, we consider individuals receiving at least one dose in some specifications reported in the robustness tests.} 
To do so, we average our dependent variable across the four weeks after school opening in every census cell $i$, and take the first-difference of this new variable across the two school years. This is done to avoid zero-inflated regressions given the high level of granularity of our dataset, and the fact that the post-summer period brings to few cases in many census areas. Defining $k$ as the week of school opening we have, therefore, $Y_{i}=\frac{\sum_{k=1}^{4}{\Delta lnCovid}_{i,k, 2021/22}}{4} -\frac{\sum_{k=1}^{4}{\Delta lnCovid}_{i,k, 2020/21}}{4}$. 

The new model, having $Y_i$ as dependent variable, takes the following form:
\begin{equation}
	Y_{i}=\alpha_i+\nu_1 \Delta Exp\_Vaccines_{age=12-19, i}  +\nu_2 \Delta Exp\_Vaccines_{age>19, i} +\theta Y_{i, k-4}+ \zeta Prov_{p}+\omega_i,
	\label{eq:np_2}
\end{equation}
where $Exp\_Vaccines_{age=12-19, i}$   and  $Exp\_Vaccines_{age>19, i}$ are two indicators of the exposure to vaccines for individuals of age between 12-19 and above 19 at census area level, built following the procedure introduced in Section \ref{se:data}. Given that vaccination was implemented only from December 27th 2020, these shares are equal to zero for the school year 2021/20, making  $\Delta Exp\_Vaccines_{age=12-19, i} =Exp\_Vaccines_{age=12-19, i}  $ and $\Delta Exp\_Vaccines_{age>19, i}=Exp\_Vaccines_{age>19, i}$, respectively. The term $Y_{i, k-4}$ denotes the lag of the dependent variable measured the four weeks before the school opening. Additionally, the control variables also include a set of municipality-level dummies to capture residual local unobserved characteristics (we exclude them in the initial specification because the municipality dummies are dropped by differencing). Equation (\ref{eq:np_2}) can be considered as a first-difference version of Equation (\ref{eq:np_1}) where the dependent variable is averaged across four weeks, to avoid zero inflated regression, and where the census area fixed effects and the week dummies are canceled out. In this framework, we still account for the initial conditions of the Covid-19 process at school openings, to control for possible differences in the phase of the epidemic between 2021/22 and 2020/21. Finally, $\omega_i$  is the error term clustered at the census area level. 

%
%


\section{Econometric Analysis}\label{se:econometrics}

In this section we present the results of the estimation of Equations (\ref{eq:np_1}) and (\ref{eq:np_2}). In particular, Section \ref{sec:bench} illustrates our benchmark findings, while Section \ref{se:rob} contains the results on the identification of heterogeneous effects and of robustness tests. Section \ref{sec:bench} also contains the results of a counterfactual analysis in which we estimate the effect of vaccinations on the number of of Covid-19 cases and on ICU beds occupancy by Covid-19 patients.

\subsection{Benchmark results\label{sec:bench}}
Table \ref{tab:1} reports the results of the estimation of Equation (\ref{eq:np_1}). In particular, Columns (1) and (2) show the coefficients from Equation (\ref{eq:np_1}) when the two school years are kept separated. The results show that the two coefficients of school opening diverge substantially in magnitude and significance level in the two school years. Specifically, the coefficient measuring the impact of school opening in 2020 on Covid-19 diffusion is positive, significant and in line with the literature also in terms of magnitude. Compared to \cite{amodio2022schools}, the coefficient is slightly lower as the time windows under consideration here are shorter.\footnote{The consideration of other time windows is discussed in Section \ref{se:rob}.}

The estimated coefficient for school opening in 2021 is not significantly different from zero, indicating that school opening is not associated to an increase in the growth rate of Covid-19 cases.\footnote{While in principle school opening may be considered as endogenous, in practice in 2020 it corresponded to a staggered design due to a national referendum for which some schools were used as polling stations (see \citealp{amodio2022schools} for details), while in 2021 a unique equal opening date was fixed for all schools, irrespectively of the number of cases or intensive care occupancy.} This finding suggests that school opening after vaccination was made available did not affect the diffusion of Covid-19, pointing to the fact that a restriction such as school closure may be not justified under these conditions. This result is confirmed when we estimate the specification on the full sample. Resulst in Column (3) show that the coefficient for school opening remains positive and significant, while the coefficient of the interaction between school opening and the year dummy for 2021 is negative, generating a null overall effect. Once again, this provides an indication that school opening in 2021 had no significant effects on Covid-19 cases diffusion, as also confirmed by the t-test on the equality of the absolute value of two coefficients reported at the bottom of Column (3). 

Finally, we slightly modify the baseline specification by adding an indicator on the number of vaccinated individuals (with second dose) of school age calculated at the time of school opening and at municipality level. Since this indicator is not time-varying, we just create a new interaction term by multiplying it with the school opening in 2021 variable. The coefficient reported in Column (4) suggests a strongly significant negative effect of this new interaction term, with less school-opening induced cases in areas where the number of vaccinated individuals was higher. Overall, these results suggest that vaccination of individuals age 12-19 may significantly contribute to the explanation of the absence of impact of school opening on Covid-19 diffusion at the beginning of the 2021/22 school year.

\begin{table}[htb]
	\singlespacing
	\captionsetup{format=plain, labelfont=bf}
	\caption{Covid-19 cases, school opening and vaccines \label{tab:1}}
	\centering
	\begin{adjustbox}{width=1\textwidth}
		\begin{threeparttable}
		\begin{tabular}{lcccc}
			\toprule
			& \multicolumn{4}{c}{Dep. Var.: Change ln of Covid-19}                                                        \\\hline
			& (1) & (2) & (3) & (4) \\ \hline
			\midrule
			Specification                             & Year=2020/21     & Year=2021/22      & Full sample  & Full Sample   \\
			\hline\\
			School opening               & \textbf{0.013***} & \textbf{0.001}   & \textbf{0.015***}    &  \textbf{0.015***}                                 \\
			& (0.002)  &   (0.002)  & (0.001)     & (0.002)                           \\
			
			School opening  X Year=2021  &          &            & \textbf{-0.016***}   & \textbf{-0.005}                                                               \\
			&          &           & (0.002)     & (0.005)                                                           \\
			
			School opening  X Year=2021/22 X Share Vaccinated (12-19)  &          &           &   &   \textbf{-0.001**}  \\
			&          &           &     & (0.001)                                        \\  
			
			&          &           &     &                                      \\  
			
			$\Delta$ ln Covid (lag)                & -0.561*** &   -0.436***        &  -0.454***           &                                              -0.454***   \\
			& (0.005)  &  (0.003)         &  (0.002)           &                     (0.002)                        \\
			ln Covid (t-3)                & -0.038***   &   -0.077***        & -0.099***            &                                              -0.099***\\
			& (0.011)  &   (0.004)        &   (0.003)          &               (0.003)                 \\
			ln Covid (t-4)                & -0.008    &     -0.011***      & -0.039***            &            -0.039***                                      \\
			& (0.011)  &   (0.004)        &  (0.003)           &                 (0.003)          \\

			\hline\\
			
			Test on school opening (prob.$>$0) & & & & \\
			$\lambda_{2020/21}+\lambda_{2021/22}$=0&-&-&0.23&- \\ \hline \\

			Census area FE                            & Y        & Y         & Y           & Y                                          \\
			Week dummies & Y        & Y         & Y           & Y                                                           \\
			Year of lag var.                          & 2020   &  2021         & 2020 $\&$ 2021        &   2020 $\&$ 2021        \\
			Observations                              & 268,832   & 268,832    & 537,664      &  537,664                                 \\
			Number of Census areas                    & 33,604    & 33,604     & 33,604       & 33,604                         \\
			$R^2$ within                                 & 0.22    & 0.22     & 0.21       & 0.21                               \\
			\bottomrule                                 
		\end{tabular}
		\begin{tablenotes}
			\item \textit{Notes: }{The table displays the results from an OLS with fixed-effect at census area level and weekly dummies. The dependent variable is the change in log of the weekly Covid-19 cases. The year 2021 dummy is added only in the interaction as it would be collinear with the week dummies. The indicator \textit{Vaccinated 12-19 (ln)} denotes the natural log of fully vaccinated individuals in school age (12-19 y.o) at the time of school opening at municipality level. This indicator is time invariant and it enters in the specification only in the interaction.  ***, **, and * denote significance at 1\%, 5\%, and 10\%, respectively. Standard errors are clustered at census area level.}{\footnotesize\par}
		\end{tablenotes}
	\end{threeparttable}
	\end{adjustbox}
\end{table}

Table \ref{tab:3} reports the coefficients obtained from the estimation of Equation (\ref{eq:np_2}), where the effect of school-age vaccination is correlated to the average growth rate of Covid-19 cases in a window of four weeks after the school opening. Column (1) shows that school-age vaccination, measured at cell level through the exposure variable (second dose), is associated to a strongly significant negative effect. This effect is robust to  the inclusion of: i) the initial conditions, i.e. the value of the average number of cases in the four weeks preceding the school opening (Column (2)); ii) the inclusion of Municipality FE (Column (3)); iii) vaccination exposure in the population older than 19 years (Column (4)). Indeed, the coefficient related to vaccination of school-age population increases in magnitude with these controls. In terms of magnitude, the coefficient on the exposure reported in Column (4) suggests that a 1\% increase on the average level of exposure to vaccination of individuals aged 12-19 is associated to a decrease of about 0.14\% in the growth rate of Covid-19 cases.\footnote{It is necessary to specify that, on average, the 12-19 population represents approximately 10\% of the cell population, therefore the exposure variable is bounded upwards.}

\begin{table}[htb]
	\singlespacing
	\captionsetup{format=plain, font=small, labelfont=bf}
	\caption{Covid-19 cases in different schooling years years and the role of vaccines \label{tab:3}}
	\centering
	\begin{adjustbox}{width=1\textwidth}
		\begin{threeparttable}
			\begin{tabular}{lcccc}
				\toprule
				& \multicolumn{4}{c}{\begin{tabular}[c]{@{}c@{}}Dep. Var.:  $Y_{i} =$ Change in the growth rate of Covid-19 cases \\  post school opening (2021-2020)\end{tabular}} \\
				\midrule
				& (1)                     & (2)                     & (3)     & (4)                \\ 
				\cline{2-5} \\
				\textbf{Exposure to vaccines (age=12-19)} & \textbf{-0.079***} & \textbf{-0.087***}  & \textbf{-0.132***}  & \textbf{-0.141***}              \\
				& (0.015)  & (0.015) & (0.015) & (0.015)    \\
				\textbf{Exposure to vaccines (age$>$19)} & & & & \textbf{-0.011***} \\
				& & & & (0.004) \\
				Change in the growth rate of Covid-19 cases \textbf{pre-school opening}                  &   & -0.062***    & -0.055*** & -0.055***                \\
				&  & (0.011)  & (0.010)  & (0.010)               \\
				
				
				\midrule

				Observations                   & 32,490                   & 32,490                   & 32,490 & 32,490                  \\
				$R^2$                    & 0.001                   & 0.004                   & 0.085 & 0.085                  \\
				Municipality FE                & N                       & N                       & Y  & Y \\
				\bottomrule                   
			\end{tabular}
			
			\begin{tablenotes}
				\item \textit{Notes:}{ ***, **, and * denote significance at 1\%, 5\%, and 10\%, respectively. Standard errors are clustered at census area level.}{\footnotesize\par}
			\end{tablenotes}
		\end{threeparttable}
	\end{adjustbox}
\end{table}
	
As an implication of the results in Table \ref{tab:3} in what follows we propose an estimation of the reduction in Covid-19 cases and of hospitalizations in ICU for Covid-19 patients, implied by the vaccination of the school-age population. Specifically, we compute the cumulated value of the cases we would have observed had vaccinations not be administered, by using the estimated coefficients from Column (4) of Table \ref{tab:3}, and setting to zero the values of school-age vaccination shares.

We first build the counterfactual difference between Covid-19 cases's growth rates in 2020 and 2021 with school age shares of vaccinated individuals equal to zero. Then we replace this difference to the Covid growth rate of cases in 2020 to build the counterfactual growth rate of 2021. Finally, starting from the level of the Covid cases 2021 before school openings, we derive the cumulated cases of 2021 4 weeks after school opening. This exercise shows that, without accounting for the vaccinations, the new weekly cases at municipality level would have increased on average by 11\% (407 vs 362).

\begin{figure}[H]
	\centering
	\minipage{0.5\textwidth} 
	\includegraphics[width=1\linewidth]{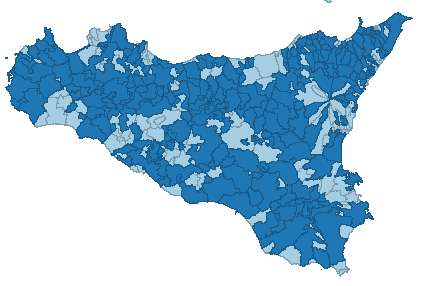}
	\endminipage\hfill
	\minipage{0.5\textwidth}
	\includegraphics[width=1\linewidth]{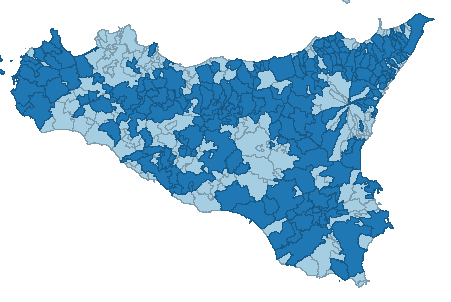}
	\endminipage\hfill
	\caption{Municipalities with more than 20 cases per 1000 inhabitants: counterfactual (left panel) and actual (right panel) scenarios}. 
	\label{Fig_scenarios}
\end{figure}

Figure \ref{Fig_scenarios} reports the comparison of the municipality average cases in the counterfactual (left panel) and actual (right panel) scenarios. Dark blue represents municipalities with more than 20 new cases per 1000 inhabitants. On average, the number of cases for 1000 inhabitants increases from 26.98 to 29.07. This seems to be a realistic result given that the share of population aged 12-19 in Sicily is 8.3\% of total population, even if it is likely to have higher interactions compared to older cohorts.

Finally, we can forecast some general effects of this vaccination. The rule for imposing restrictions on economic activities, schools, etc. in Italy (by imposing the so called ``yellow" and ``red zones", implying respectively some minor restrictions as masks in open areas, and major restrictions such as school closures and limitations on mobility) assumes among other parameters a municipality average ICU occupancy by Covid-19 patients respectively of 10\% and 30\%. 

The average municipality number of ICU beds in Sicily is 1.02 but this number is driven by small municipalities without ICU beds, otherwise the average number would be 13.8.\footnote{The average distance of a municipality without ICU beds from the nearest municipality with ICU beds is 3.1 km.} Then, in a very conservative scenario for this smaller set of municipalities with ICU beds, this means reaching the critical ICU parameter for the declaration a status of ``yellow zone" and ``red zone" respectively with the second and fifth occupied ICU bed by a Covid-19 patient.
 
To build a counterfactual scenario in terms of ICU beds occupancy, consider that the ratio between active cases and ICU patients was 236.5 in Sicily in the four weeks after the 2021 school openings of September 16th. This implies that the additional cases implied by our counterfactual scenario with no school-age vaccinations would have implied a 19.03\% increase of ICU bed occupancy by Covid-19 patients, a remarkably high number.


\subsection{Heterogeneity and Robustness \label{se:rob}}
\begin{figure}[H] 
\label{Fig_het}
	\begin{centering}
		\includegraphics[width=0.80\textwidth]{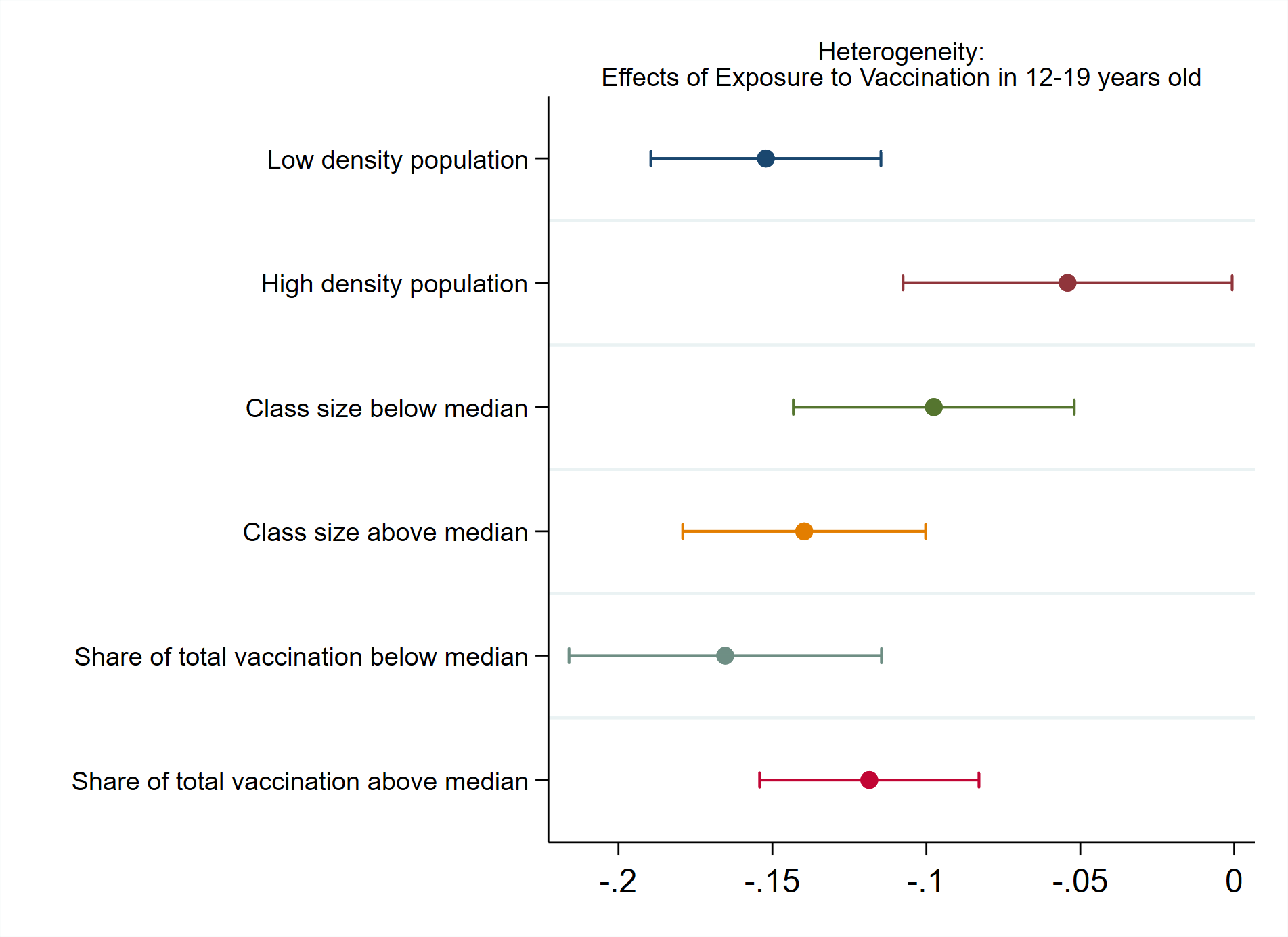}
		\caption{Heterogeneity: Effects of Exposure to Vaccination in 12-19 years old}
		\label{fig:het} 
	\end{centering}

\end{figure}

We carried out a set of exercises to test whether the relationship between Covid-19 cases, vaccination and school opening may be heterogeneous across a set of socio-demographic characteristics that may be linked to Covid-19 diffusion. Specifically, we test for heterogeneity across population density, classroom size and vaccination share of the whole population, by estimating Equation (\ref{eq:np_2}) on two subsamples deriving from splitting the sample by the median value of these dimensions.\footnote{We obtained similar results when splitting the sample above and below the mean values. Results are available upon request.} Figure \ref{fig:het} reports the whisker plot of the coefficients obtained from this exercise, while Table \ref{tab:app3} in the Appendix reports the full sets of coefficients.

As Figure \ref{fig:het} shows, the coefficients of the exposure variables are all significant and negative, and the confindence intervals for most of them largely overlap. Interestingly, the only dimension for which the two coefficients do not largely overlap is the density of population, suggesting that the effect of vaccination exposure is larger in absolute terms for areas with lower population density. 

This result is similar to \cite{amodio2022schools}, who find a larger effect of the 2020 school opening in low population density  zones. This finding is likely due to the higher degree of social interactions in smaller sized communities as pointed out in \cite{sato2015urbanization}.

Finally, we test the robustness of our main results through a set of additional exercises, that we compare to those of our preferred specification Column (3) of Table \ref{tab:1}, which are reported in Column (1) of Table \ref{app:tablerob}, which contains the results of the robustness tests.  First, we test for eventual bias stemming from the inclusion of the lagged dependent variable in the dynamic process of Equation (\ref{eq:np_1}). This is done following the work of \cite{Chernozhukove2103420118}, which uses the Analytical Bias Correction estimator of \cite{chen2019mastering} to take into account the potential Nickell bias in this process. Column Column (2) of Table \ref{app:tablerob} shows that results are substantially unchanged. Then, we check whether the definition of the time-window may affect the results. We therefore extend the time window up to 8 weeks before and after the school opening and run the baseline specification, without not finding any appreciable difference (Column (3) of Table \ref{app:tablerob}. We then take into account potential spillover effects due by the school opening, by using the process of \cite{conley1999gmm}, implemented by \cite{colella2020acreg}. Column (4) of Table \ref{app:tablerob}) shows that the coefficients remain consistent. Finally, we model school opening in 2020 through propensity scores, to take into account the small flexibility implied by the referendum, following the same procedure of \cite{amodio2022schools}. Column (5) of Table \ref{app:tablerob} shows that he results remain robust when accounting for this potential source of endogeneity (see \citealp{amodio2022schools}, for more details on the use of propensity scores). 

We also carried out a set of robustness tests on the specification of Equation (\ref{eq:np_2}), where we modified the definition of the dependent variable and used the share of population that received at least a single dose of vaccine. As reported in Table \ref{app:tablerobexp}, whose Column (1) contains our benchmark results, i.e. those in Column (4) of Table \ref{tab:3}, the coefficients appear slightly more variable in Columns (2) and (4), but this is likely to depend on the change in the scale of the dependent and independent variables, which are hereby computed differently from before. Specifically, in Column (2) of Table \ref{app:tablerobexp} we consider an alternative specification in which we rearrange the time span of the regression as follows: (1) the dependent variable is calculated as the growth rate of Covid-19 cases in the 4 weeks after school reopening in 2020 and 2021, and (2) the control variable is given by the level of Covid-19 cases in the 4 weeks before school opening in 2020. Differently, in Column (4) of Table \ref{app:tablerobexp} we consider at least the first vaccination shot to build the exposure variable, whose average is 2.3 times bigger than in the benchmark case, so that if we apply this change to the coefficient we obtain -0.06, which is less than half of the benchmark value (-0.14) but it remains negative and significant.

\section{Conclusions}\label{se:conclusions}
The Covid-19 pandemic hit all countries as an unexpected shock in 2020 when few pharmaceutical tools were present. This implied that the first set of governmental reactions was mainly based on restrictions on individuals' mobility to lower the frequency of interaction and the consequent spread of the virus. Subsequently, medical treatments, mostly in the forms of vaccines, allowed to increase and improve the set of tools to contrast the Covid-19 diffusion, which started to imply a trade/off in using the restrictions. The evidence presented in this paper shows that school age vaccination had a substantial role in reducing, basically neutralizing, the effect of school openings on Covid-19 diffusion. With a focus on Sicily and on a rich set of granular data, while school openings were a substantial driver of cases in 2020, this effect disappears in 2021, when school-age vaccination was available.

In particular, our results show that an additional 1\% of vaccination in the age cohort 12-19 (attending middle and high school) is associated to a decrease in 0.14\% of Covid growth rate in post-summer school reopening period at local level. In addition, in a counterfactual analysis we show that school-age vaccination is associated to an estimated reduction of 19\% occupancy of ICU beds by Covid-19 patients, which implies a sizeable effect on the possibility of municipalities to escape the restrictions which would be otherwise implemented by the State.

Several points are left to future agenda, in which we can consider how they may make our results appear to be a lower bound or an upper one. Specifically, in this paper we assumed that attitudes of vaccinated individuals do not change,\footnote{If the evidence of \cite{andersson2021anticipation} holds, we should expect vaccinated individuals to implement lower social distances than individuals not vaccinated so that, ceteris paribus, the effects of vaccination should have been bigger.}while individual restrictions affecting unvaccinated individuals in Italy (even if only for people older than 17, a small share of our class age) may imply that vaccinated individuals have higher mobility, so that the mitigation effect caused by vaccination should be counterfactually lower. Then, we could take into account that the school opening period is a post-summer period so that, due to the seasonality, we would observe less cases. This means that our results could be a lower bound, because in a period of larger spread of diffusion this effect could have been larger.

\pagebreak
\clearpage
\bibliography{covid}
\vfill
\clearpage

\appendix
\renewcommand*\appendixpagename{Appendix}
\renewcommand*\appendixtocname{Appendix}
\appendixpage
\addappheadtotoc

\section{Additional Results}
In this Appendix, we present some additional results. Table \ref{tab:app3} contains the results on heterogeneity of the effects estimated from Equation (\ref{eq:np_1}). Table \ref{app:tablerob} contains the results of the robustness tests for the results from the estimation, respectively, of Equations (\ref{eq:np_1}) and (\ref{eq:np_2}).

\setcounter{table}{0}
\begin{table}[htb]
	\renewcommand{\thetable}{\thesection\arabic{table}}
	\singlespacing
	\captionsetup{format=plain, font=small, labelfont=bf}
	\caption{Heterogeneity: Covid-19 cases through the years and the role of vaccines \label{tab:app3}}
	\centering
	\begin{adjustbox}{width=1\textwidth}
		\begin{threeparttable}
			\begin{tabular}{lcccccc}
				\toprule
				& \multicolumn{6}{c}{\begin{tabular}[c]{@{}c@{}}Dep. Var.: Yi = Change   in the growth rate of Covid-19 cases\\      post school opening (2021-2020)\end{tabular}} \\
				\midrule
				& \begin{tabular}[c]{@{}c@{}}Low density \\ population\end{tabular} & \begin{tabular}[c]{@{}c@{}}High density \\ population\end{tabular} & \begin{tabular}[c]{@{}c@{}}Class-size \\ $<$ median\end{tabular} & \begin{tabular}[c]{@{}c@{}}Class-size \\ $>$ median\end{tabular} & \begin{tabular}[c]{@{}c@{}}Share of total \\ vaccination $<$ median\end{tabular} & \begin{tabular}[c]{@{}c@{}}Share of total \\ vaccination $>$ median\end{tabular} \\
				\cline{2-7}
				Exposure to vaccines (age=12-19) & -0.152*** & -0.054** & -0.098*** & -0.140*** & -0.165*** & -0.119*** \\
				& (0.019) & (0.027) & (0.023) & (0.019) & (0.026) & (0.018) \\
				Exposure to vaccines (age\textgreater{}19) & -0.007* & 0.003 & -0.007 & -0.011** & -0.015** & -0.007 \\
				& (0.004) & (0.010) & (0.007) & (0.004) & (0.006) & (0.004) \\
				\begin{tabular}[c]{@{}l@{}}Change in the growth rate of \\ Covid-19 cases  pre-school opening\end{tabular} & -0.053*** & -0.050*** & -0.063*** & -0.047*** & -0.079*** & -0.028* \\
				& (0.017) & (0.012) & (0.016) & (0.013) & (0.014) & (0.014) \\
				\midrule
				Observations & 15992 & 16538 & 11635 & 21010 & 16460 & 17011 \\
				$R^2$ & 0.091 & 0.144 & 0.140 & 0.085 & 0.101 & 0.071 \\
				Municipality FE & Y & Y & Y & Y & Y & Y \\
				\bottomrule
			\end{tabular}
			\begin{tablenotes}
				\item \textit{Notes: }{ ***, **, and * denote significance at 1\%, 5\%, and 10\%, respectively. Standard errors are clustered at census area level.}{\footnotesize\par}
			\end{tablenotes}
		\end{threeparttable}
	\end{adjustbox}
\end{table}

\begin{table}[htb]
	\renewcommand{\thetable}{\thesection\arabic{table}}
	\singlespacing
	\captionsetup{format=plain, labelfont=bf}
	\caption{Robustness tests}
	\label{app:tablerob}
	\centering
	\begin{adjustbox}{width=0.9\textwidth}
		\begin{threeparttable}
			\begin{tabular}{lccccc}
				\toprule
				& \multicolumn{5}{c}{Estimation method}                                                        \\\hline
				& (1) & (2) & (3) & (4) & (5) \\\hline
				\midrule                             & Benchmark&ABC est.     & Longer time window       & AC corr. & PS   \\
				\hline\\
				School opening               & \textbf{0.015***}   & \textbf{0.012***}    &\textbf{0.015***} &\textbf{0.014***}&                           \textbf{0.016***} \\
				& (0.001)  &                            (0.001) & (0.001) &(0.003)&  (0.003)  \\

				School opening X Year=2021/22 &  \textbf{ -0.016***}       & \textbf{ -0.014***} &\textbf{ -0.016***} &\textbf{ -0.016***}& \textbf{-0.022***}  \\
				&  (0.002)          &     (0.002)        &    (0.002)  &(0.004)&           (0.003)                                               \\

				\\ \hline \\

				Other Controls                           & Y        & Y         & Y  &Y&   Y                                       \\
				Census area FE                            & Y        & Y         & Y &Y& Y                                    \\
				Week dummies       & Y         & Y           & Y          &Y&    Y                                             \\
				Observations                              & 537,664  &537,664& 1,041,724               & 530,944 & 530,944     \\
			Census areas                    & 33,604    & 33,604     & 33,604   &33,184&  33,184  \\
				$R^2$ within                                 & 0.21    & -     & 0.21 &0.21&        0.21      \\
				\bottomrule                                 
			\end{tabular}
		\begin{tablenotes}
			\item \textit{Notes: }{The table displays the results from an Analytical Biased corrected diff-in-diff estimation of the specifications in Table 2.  ***, **, and * denote significance at 1\%, 5\%, and 10\%, respectively. Standard errors are clustered at census area level.}{\footnotesize\par}
		\end{tablenotes}
		\end{threeparttable}
	\end{adjustbox}
\end{table}

\begin{table}[htb]
	\renewcommand{\thetable}{\thesection\arabic{table}}
	\singlespacing
	\captionsetup{format=plain, labelfont=bf}
	\caption{Robustness test on the exposure specification}
	\label{app:tablerobexp}
	
	\centering
	\begin{adjustbox}{width=0.9\textwidth}
		\begin{threeparttable}
			\begin{tabular}{lcccc}
				\toprule
				& \multicolumn{4}{c}{Estimation method}                                                        \\ \hline
				& (1) & (2) & (3) & (4) \\\hline
				\midrule                             & Benchmark&Alt. est.     & Longer sample       & $1_{st}$ dose   \\
				\hline\\
				\textbf{Exposure to vaccines (age=12-19)}               & \textbf{-0.141***}   & \textbf{-0.678***}    &\textbf{-0.105***} &\textbf{-0.059***}\\
				& (0.015)  &                            (0.061) & (0.008) &  (0.006)  \\
				
				
				\\ \hline \\
				
				Other Controls                           & Y        & Y         & Y  &   Y                                       \\
				Census area FE                            & Y        & Y         & Y & Y                                    \\
				Week dummies       & Y         & Y           & Y          &    Y                                             \\
				Census areas                    & 32,490    & 33,184     & 32,490   & 32,490    \\
				$R^2$ within                                 & 0.09         & 0.21 &0.14& 0.09             \\
				\bottomrule                                 
			\end{tabular}
			\begin{tablenotes}
				\item \textit{Notes: }{***, **, and * denote significance at 1\%, 5\%, and 10\%, respectively. Standard errors are clustered at census area level.}{\footnotesize\par}
			\end{tablenotes}
		\end{threeparttable}
	\end{adjustbox}
\end{table}

\end{document}